# ABR Flow Control for Multipoint Connections

Sonia Fahmy and Raj Jain

Department of Computer and Information Science

The Ohio State University

2015 Neil Avenue, DL 395

Columbus, OH 43210-1277, USA

Tel: (614) 688-4482, Fax: (614) 292-2911

E-mail: {fahmy, jain}@cis.ohio-state.edu

## ABR Flow Control for Multipoint Connections

Multipoint capabilities are essential for ATM networks to efficiently support many applications, including IP multicasting and overlay applications. The current signaling and routing specifications for ATM define point-to-multipoint capabilities. Multipoint-to-point connection support is also being discussed by the signaling and PNNI groups, and will be defined in the near future for the unspecified bit rate (UBR) service. Below, we examine point-to-multipoint and multipoint-to-point flow control for the available bit rate (ABR) service, as discussed in the traffic management working group.

The ABR service attempts to provide possibly non-zero minimum rate guarantees, achieve fairness, and minimize cell loss for data (non real-time) traffic, by periodically indicating to ABR sources the rates at which they should be transmitting. The service uses closed-loop end-to-end feedback control. The feedback from the switches to the sources is indicated in resource management (RM) cells generated by the sources and turned around by the destinations. The RM cells flowing from the source to the destination are called forward RM cells (FRMs) while those returning from the destination to the source are called backward RM cells (BRMs). The RM cells contain the source current cell rate (CCR), in addition to several fields that can be used by the switches to provide feedback to the sources. Feedback can be just one or two bits to indicate congestion, and/or the exact rate at which the source should transmit, called the explicit rate (ER). When a source receives a BRM cell, it computes its allowed cell rate (ACR) using its current ACR value, the congestion indication bits, and the ER field of the RM cell [5].

For ABR point-to-multipoint connections, branch points replicate data and FRM cells, and consolidate feedback information from the different branches. Feedback consolidation can be explained with the aid of figure 1. The consolidation operation is required to avoid the feedback implosion problem, where the number of BRM cells received by the source is proportional to the number of leaves in the multicast tree. In addition, the source ACR should not keep fluctuating due to the varying feedback received from different leaves. Consolidation noise can occur when feedback from some branches is not always received at the time when the RM cells need to be returned by the branch point.

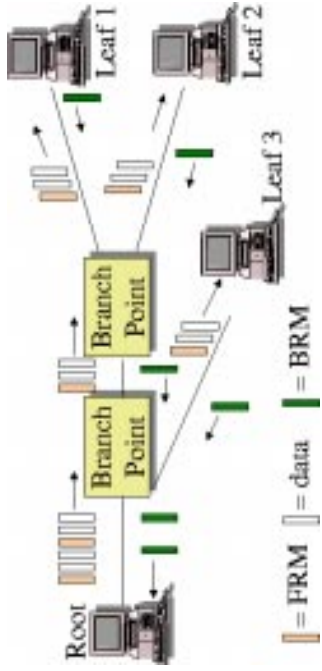

Figure 1: Point-to-multipoint connections

In *point-to-point* ABR flow control, the source is controlled to the minimum rate that can be supported by all the switches on the path from the source to the destination [5]. This strategy can be extended to *point-to-multipoint* connections by controlling the source to the minimum rate that can be supported by the switches on the paths from the source to *all of the leaves* in the multicast tree. This is because the minimum rate is the technique most compatible with the typical data requirements: no data should be lost, and the network can take whatever time it requires to deliver the data intact [11].

The BRM consolidation method at the branch points should: (1) Preserve the efficiency and fairness properties of the rate allocation schemes employed in the network switches, (2) Scale well with the number of levels and with the number of branches in the multicast tree, (3) Maintain the ratio of BRM cells to FRM cells in the network and at the root close to one during normal operation, (4) Exhibit a reasonable transient response, and handle non-responsive branches such that they do not halt the consolidation operation nor cause overload or underload, and (5) Exhibit little consolidation noise and consolidation delays.

There are several ways to implement the consolidation algorithm at branch points [4, 7, 8, 11, 13]. Each method offers a tradeoff in complexity, scalability, overhead, transient response, and consolidation noise. These tradeoffs can be summarized as follows [4]:

(1) Which component generates the BRM cells (i.e., turns around the FRM cells)? Should the branch point, or should the destinations (leaves), perform this operation?

(2) Should the branch point wait for feedback from all the branches before passing the BRM cell upstream? Although this alleviates the consolidation noise, it may incur additional complexity or increase the transient response of the scheme, especially after idle or low rate periods. (The slow transient response, however, can be mitigated in cases of severe overload where the algorithm can immediately indicate the overload information to the source [4].)

(3) How can the ratio of FRM cells generated by the source to BRM cells returned to the source be controlled?

(4) How can the ratio of BRM cells in the network to the source-generated FRM cells be controlled?

(5) How does the branch point operate when it is also a switch (i.e., a queuing point)? The coupling of the rate allocation and branch point functions must be considered. When should the actual rate computation algorithm be invoked?

(6) How can the scheme scale well to large multicast trees? Will the feedback delay grow with the number of levels of the tree?

(7) How is accounting performed at the branch point? Consolidation algorithms use registers to store values such as the minimum rate given by branches in the current iteration, and flags to indicate whether an RM cell has been received since the last one was sent. Some values should be stored per branch, while others should be maintained per connection, regardless of the number of branches.

(8) How are non-responsive branches handled? If the consolidation scheme waits for feedback from all the branches before sending a BRM to the source, an algorithm must be developed to determine when a branch becomes non-responsive.

So far, we have discussed the issues with point-to-multipoint ABR connections. Although *multipoint-to-point ABR* connections are not being discussed at the control signaling and PNNI working groups yet [12], some preliminary work has been done on the merge point operation [3, 6, 10, 9]. Merge points must ensure that BRM cells are sent to the appropriate sources at the appropriate times (see figure 2). The feedback regulation algorithm should be simple, scalable, and minimize noise and delays. It should also maintain the BRM to FRM ratio at the sender and inside the network close to one during normal operation.

For multipoint-to-point connections, the assumption implicit in some ABR rate allocation mechanisms that each connection has only one source is no longer valid. Suppose we consider the traffic of every virtual connection (VC) coming on an input link to a merge point as constituting a separate *flow*. Then, multiple flows can be merged into *one* flow if they originate from sources coming from different input links, but belonging to the same VC. Thus, four different types of fairness can be defined for multipoint-to-point VCs [3]: (1) **Source-based fairness,** which divides bandwidth fairly among active sources as if they were sources in point-to-point connections, ignoring group memberships, (2) **VC/source-based fairness,** which first gives fair bandwidth allocations at the VC level, and then fairly allocates the bandwidth of each VC among the active sources in this VC, (3) **Flow-based fairness,** which gives fair allocations for each active flow, (4) **VC/flow-based fairness,** which first divides the available bandwidth fairly among the active VCs, and then divides the VC bandwidth fairly among the active flows in the VC.

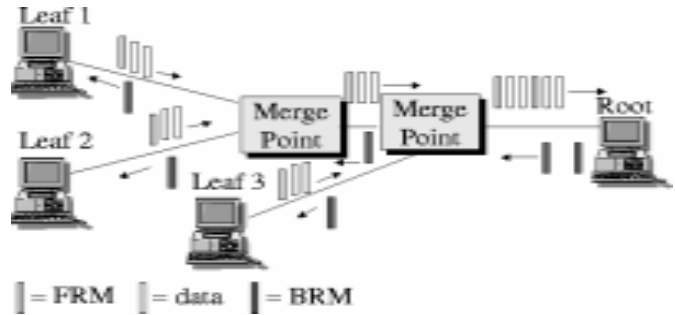

Figure 2: Multipoint-to-point connections

Rate allocation algorithms for multiple sender connections must avoid performing source-level accounting. For example, measuring the rates or activity for each source, or distinguishing overloading and underloading sources cannot be performed. This is because cells from different sources in the same VC may be indistinguishable after merging in some switches. If accounting is performed at the VC level or at the flow level, an additional mechanism to divide VC or flow bandwidth among sources is necessary. In addition, CCR values from BRM cells should not be used in computing rate allocations for sources in multipoint-to-point connections, since the CCR value in the BRM can be that of a source whose traffic does not go through the merge point.

Example branch point algorithms are given in the current baseline text [1], while example merge point algorithms are given in the living list of the traffic management working group [2]. Multipoint-to-multipoint connections can be handled by combining a point-to-multipoint (branch point) algorithm with a multipoint-to-point (merge point) algorithm [10]. This will be the subject of future studies.

**Sonia Fahmy [StM]** received her MS degree in Computer Science in 1996 from the Ohio State University, where she is currently a PhD student. Her main research interests are in the areas of multipoint communication, traffic management, and performance analysis. She is the author of several papers and ATM Forum contributions. Internet: http://www.cis.ohio-state.edu/~fahmy/

**Raj Jain [F]** is a Professor of Computer and Information Science at the Ohio State University in Columbus, Ohio. He is the author of "The Art of Computer Systems Performance Analysis," published by Wiley in 1991. Dr. Jain is an IEEE Fellow, an ACM Fellow and is on the editorial boards of several journals. Internet: http://www.cis.ohio-state.edu/~jain/